\documentstyle[aps,prb,epsf]{revtex}

\def\be{\begin{equation}}
\def\ee{\end{equation}}
\def\bea{\begin{eqnarray}}
\def\eea{\end{eqnarray}}

\begin{document}

\title{QUANTIZED VORTICES IN SUPERFLUIDS AND SUPERCONDUCTORS}

\author{D. J. THOULESS$^1$, Ping AO$^{1,2}$, Qian NIU$^{3}$,
M. R. GELLER$^{1,4,5}$ and C. WEXLER$^{1,6}$}

\address{ $^1$University of Washington, Seattle, WA, USA\\
e-mail thouless@phys.washington.edu\\
$^2$Ume\aa \ University, Sweden\\
$^3$University of Texas, Austin, TX, USA\\
$^4$Simon Fraser University, Burnaby, Canada\\
$^5${\rm Present address:} University of Georgia, Athens, GA, USA\\
$^6${\rm Present address:} University of Florida, Gainesville, FL, USA}


\maketitle

\begin{abstract}{ We give a general review of recent developments
in the theory of vortices in superfluids and superconductors,
discussing why the dynamics of vortices is important, and why some key
results are still controversial.  We discuss work that we have done on
the dynamics of quantized vortices in a superfluid.  Despite the fact
that this problem has been recognized as important for forty years,
there is still a lot of controversy about the forces on and masses of
quantized vortices.  We think that one can get unambiguous answers by
considering a broken symmetry state that consists of one vortex in an
infinite ideal system.  We argue for a Magnus force that is
proportional to the superfluid density, and we find that the effective
mass density of a vortex in a neutral superfluid is divergent at low
frequencies.  We have generalized some of the results for a neutral
superfluid to a charged system.}
\end{abstract}

\section{Introduction}

From the very beginning it has been realised that quantized vortices play an
important part in the behavior of superfluids \cite{onsager49}.  Both in
neutral superfluids and in superconductors it is the vortices that provide a
mechanism for the decay of superfluid currents in a ring.  The circulation,
for a neutral superfluid, or the trapped flux, for a superconducting ring, is
quantized, and the current can only decay by a change of the quantum number
by an integer, which can occur by the passage of a vortex (or quantized flux
line) across the ring, from one edge to the other.  In superconductors in a
high magnetic field, the motion of flux lines is the main mechanism for
electrical resistance.  At high temperatures the movement of vortices is a
thermally activated process, but at low enough temperatures the dominant
mechanism must be by quantum tunneling.  It is therefore important to
understand the dynamics of vortices, in order to be able to evaluate the
dissipative processes that occur in neutral superfluids and in
superconductors.

Despite the obvious importance of the problem, the theory has been in a most
unsatisfactory state.  There are many conflicting results in the
literature.  We did not realise the extent of this disagreement, and it was
initially a surprise to us that a result we obtained would be dismissed by one
knowledgeable critic as too obvious to be worth discussing, and by another
equally eminent critic as well known to be wrong; this has happened to us
several times.  There are real problems here, connected with questions of
suitable boundary conditions, and there is often a question of whether two
different calculations are finding the same result by two different ways, or
if they are finding two different contributions which must be added.

To our surprise, we found, about two years ago, that we could get an exact
result for one of the two parameters that determine the transverse force on
a moving vortex, using only general properties of superfluid order
\cite{tan96}.  The second parameter was determined by a straightforward
thermodynamic argument a little over a year later \cite{wexler97}.  The first
of these results still seems to be controversial \cite{volovik96}, and some
elements of the argument deserve closer scrutiny than they have received so
far, but it is our belief that it should be possible to construct a firmly
founded theory on the basis that we have tried to establish.

\section{Electrons in magnetic fields and vortices}

There are strong analogies between the behavior of electrons in strong
magnetic fields and of vortices in superfluids.  These analogies enable us
to make use of some of the insights that have been obtained in the study of
the quantum Hall effect to understand problems connected with vortex
dynamics. 

In both cases there is a transverse force proportional to velocity. The
Lorentz force for electrons is proportional to the vector product of the
electron velocity and the magnetic field, ${\bf F}_L=-e{\bf v}\times{\bf B}$.
This can be represented by a term $eBx\dot y$ in the Lagrangian. The Magnus
force acting on a vortex is proportional to the vector product of the
velocity of the vortex relative to the fluid and a vector directed along the
vortex core.  Each of these forces leads to a path-dependent but speed
independent term in the action. In a quantum theory the phase is equal to the
action divided by $\hbar$, this corresponds to a Berry phase, a phase which
depends on the path, but not on the rate at which the path is
traversed \cite{haldanewu85,ao93}.  

In both cases there is considerable arbitrariness in the value of this phase. 
For the electron this is due to the arbitrariness of the vector
potential which is used to represent the magnetic field, while for the vortex
there is a similar arbitrariness in the way the transverse force is
represented in a Lagrangian.  In either case the change in action or phase
when the electron or vortex traverses a {\it closed} path is well
determined.  For the electron the phase change on a closed path is equal to
$2\pi$ times the number of flux quanta enclosed by the path.  For the vortex
the phase change is equal to $2\pi$ times the average number of atoms
enclosed by the surface swept out by the closed path of the vortex.

The dominant correction to quantization of the Hall conductance comes from
tunneling or activated transport between states on the two edges of a
quantum Hall bar. The dominant mechanism for decay of supercurrents is
tunneling or activated transport of vortices across the system.  In many
real systems there is a tangle of pre-existing vortices frozen in when the
system is cooled below the critical temperature, and these can serve as
sources for the vortices that cross the system.  Under ideal conditions, and
some modern experiments on helium approach such ideal
conditions \cite{avenel94,packard95}, there are no vortices in equilibrium, and
a vortex loop must be created from nothing in the interior, or a line must be
created at the boundary (with its image this constitutes a loop), and cross
the system to be annihilated at the opposite boundary.

Electrons in a magnetic field have a fast cyclotron motion around the
guiding center. Canonical variables can be chosen as the two pairs
$v_x,v_y$, rescaled by $m^2/eB$, which give the fast motion, and the guiding
center coordinates $X,Y$, rescaled by $eB$, which give the slow motion.
Vortices also have such a fast cyclotron-like motion, in which the vortex
core circles around the center of the flow it induces.  In addition, since
the vortex is a string rather than a point, it has the low frequency Thomson
modes, circularly polarized modes of oscillation analogous to the modes of
oscillation of a stretched string.  For the vortex the two coordinates $X,Y$
of the position of the vortex in the plane perpendicular to the core are
also conjugate variables, as is manifest from the classical theory of vortex
motion which can be found in Lamb's {\it Hydrodynamics}, and which Lamb
credits to an 1880's book on Mechanics by Kirchhoff.

The most important difference is that we think we understand the
Schr\"odinger equation for electrons, whereas a vortex in a superfluid is
a complicated many-body entity. The relation of the collective variables
describing the vortex motion to the single-particle variables describing the
superfluid is not obvious. \vskip 18pt

\section{General features of vortices in superfluids}

A vortex is a composite object in a many-body system.  Its motion may be
described by collective variables, but its structure depends on all the
single-particle variables of the superfluid, and the relation between these
single-particle variables and the collective variables is, as usual,
obscure. Feynman \cite{feynman55} proposed describing a vortex by taking the
ground state wave function, symmetric in all the single particle variables
for a boson superfluid, and multiplying it by a factor of the form 
\be\prod_{j=1}^N e^{i\theta_j}f(r_j)\;,\ee
where $\theta_j$ is the azimuthal angle made by the particle $j$ with
the vortex core, $r_j$ is its distance from the core, and $f$ is some
real function which is close to unity everywhere except where $r_j$ is
of the order of the radius of the vortex core, and which goes to zero at
$r_j=0$ in order to prevent large kinetic energy contributions from
the rapid variation of phase close to the vortex core.  A similar
description of the core is obtained for the Ginzburg--Pitaevskii
equations for the order parameter near the critical temperature
\cite{ginzburgpita58}, or from the Gross--Pitaevskii nonlinear
Schr\"odinger equation for the condensate of the dilute Bose gas at
zero temperature \cite{pitaevskii61,gross61}.  In these theories the
vortex is described by mean-field-like equations, so that the position
of the singularity at the vortex core has a sharp value, although we
know that the two components of its position in the plane are
conjugate variables.  Somehow we should be able to construct a
quantized version of the theory taking account of this result of the
Magnus force.

In a strongly type II (Shubnikov) superconductor the situation is somewhat
similar, except that the current circulating round the vortex core generates
a magnetic field parallel to the core, which in turn generates a vector
potential that reduces the current, so that a total of one quantum of flux
$h/2e$ is associated with the vortex, or flux line, and no current is
associated with the change of the phase angle at large distances.  In a type
I (Pippard) superconductor the character of the singularity is mainly
trapped flux, but the singly quantized flux line is not stable in a uniform
magnetic field, and it is thermodynamically favorable for the flux lines to
aggregate and form a region of normal metal.  In either case
Landau--Ginzburg theory can be used to describe the vortex.

In classical incompressible fluid mechanics the hydrodynamic mass of a
vortex is of order of mass of fluid displaced, but it depends in
detail on the core structure.  Since vortices in low temperature
superfluid helium are measured to have a rather small vortex core
radius, smaller than the average interatomic spacing, this mass
density is relatively small, and is taken to be zero in some
calculations.  In recent work Duan and Leggett showed that the
inertial mass of a vortex in a superconductor is
finite\cite{duanleggett92}, but Duan argued that the mass density of a
vortex in a neutral superfluid is infinite\cite{duan94}.  He
originally described this as a result of the quantum nature of the
fluid, and we found this very hard to accept.  Actually it is true of
all compressible fluids, but the divergence is logarithmic in the
frequency of the motion, with quite a small coefficient, as Demircan,
Ao and Niu have pointed out\cite{demircan96}.  Under realistic
circumstances, such as in the free cyclotron motion of the vortex, or
in vortex tunneling, the logarithm may be quite small, and this term
relatively unimportant.

In liquid helium at relatively high temperatures, close to the
critical temperature, the largest force on a moving vortex, or on a
vortex that is held still while the fluid streams past it, is likely
to be a drag force due to the scattering by the vortex of the
excitations that make up the normal fluid.  At lower temperatures the
transverse (Magnus) force should dominate, but the understanding of
the Magnus force is complicated by the existence of the two components
of the fluid, which may affect the vortex very differently.  As we
have discussed already, the Magnus force has important implications
for the Berry phase, and for the quantum uncertainty of the position
of the vortex.

For a superconductor the situation is far more complicated, since not
only is there the magnetic field due to the motion of the electrons to
be considered, but the effects of disorder in the positive background
are vital. Disorder makes the {\it conductivity} of the normal metal
finite, and produces a drag force on vortices even at rather low
temperatures, but also, if the disorder is on a large enough scale,
pins the vortices and reduces the flux flow {\it resistivity}.

In our work we have concentrated on understanding an isolated
vortex in an ideal, uniform, infinite superfluid.  Our aim has been to
understand the parameters that come into the dynamics of a vortex when its
velocity relative to the background fluid is small --- the effective mass,
the transverse component of the force, and the longitudinal (dissipative)
component of the force.  This is clearly not a program for a complete
understanding of vortex dynamics, since, even if it were completely
successful, we might still be concerned with strongly nonlinear regions in
realistic situations, such as those found when quantum tunneling appears to
be observed.  Particularly for the transverse force, we think we have clean
and precise results that are --- inevitably--- in conflict with widely
accepted theories.

In our work the vortex is controlled by some pinning potential that can be
manipulated from outside. The pinning potential can be rather weak, or a
macroscopic wire, so long as it has cylindrical symmetry.  For quantities
such as the effective mass and the longitudinal force the nature of the
pinning potential has an effect on the answer, and we may need to consider
some suitable limiting process to make the strength of the potential tend to
zero, but for the transverse force we find that the answer is independent
of the form or strength of the pinning potential.  \vskip 18pt

\section{The Magnus force in neutral superfluids}

There is no agreement about what the forces acting on a vortex in a
neutral superfluid are.  The simplest quantity to calculate should be the
component of force perpendicular to the motion of the vortex relative to
the substrate, the analog of the Magnus force for classical fluids, yet two
recently quoted forms look quite different.  In Donnelly's book on {\it
Quantized Vortices in Helium II} \cite{donnelly91} he quotes the force per
unit length as 
\be {\bf F}_t= {\bf K}\times\bigl[ \rho_s({\bf v}_V-{\bf v}_s)
+(\rho_n-\sigma) ({\bf v}_V-{\bf v}_n)\bigr]\;,\label{eq:donnelly}\ee
where ${\bf K}$ is a vector along the vortex line whose magnitude is
the quantum of circulation $h/M$, ${\bf
v}_V$, ${\bf v}_s$ and ${\bf v}_n$ are the velocities of the vortex, the
superfluid component and the normal fluid component, and $\rho_s$ and
$\rho_n$ are the superfluid and normal fluid densities; $\sigma$ is a
coefficient whose value is not exactly determined.  Volovik, however, in a
number of recent papers \cite{volovik93,volovik95}, quotes the form  
\be {\bf F}_t= {\bf K}\times \bigl[\rho({\bf v}_V-{\bf v}_s) +\rho_n({\bf
v}_s-{\bf v}_n)+C_F({\bf v}_n-{\bf v}_V)\bigr]\;,\label{eq:volovik}\ee  
where the term with coefficient $C_F$ occurs only for fermion superfluids,
such as the $B$ phase of superfluid $^3$He, and is due to spectral flow of
the low energy states in the vortex core.  The first term in each of these
expressions  is referred to as the {\bf Magnus force}, the term
proportional to $\rho_n$ as the {\bf Iordanskii force}, so the use of these
two terms is quite different for the two authors.  Donnelly's term
proportional to $\sigma$ comes from phonon or roton scattering by the vortex, and it
is only if this is equal to zero that the two expressions are in
agreement for the case of bosons.

Whatever the form of this force, Galilean invariance tells us that
there are only two parameters to be determined.  If we know the
coefficients of ${\bf v}_V$ and ${\bf v}_s$, the coefficient of ${\bf
v}_n$ must be equal to minus the sum of the other two coefficients.
We argue, in the rest of this section, that the only transverse force
has the form 
\be {\bf F}_t= \rho_s{\bf K}\times ({\bf v}_V-{\bf v}_s)
\;,\ee 
by determining separately the coefficients of $v_s$ and $v_V$.

Wexler \cite{wexler97} has given a thermodynamic argument to show that the
coefficient of ${\bf K}\times {\bf v}_s$ is indeed $-\rho_s$.  This result
seems to be uncontroversial, and is in agreement with both Donnelly and
Volovik.  The argument is essentially a thermodynamic argument, which
considers a reversible change of the circulation in a ring by moving a vortex
slowly across the system under equilibrium conditions.

\begin{figure}
\begin{center}
\leavevmode
\epsfbox{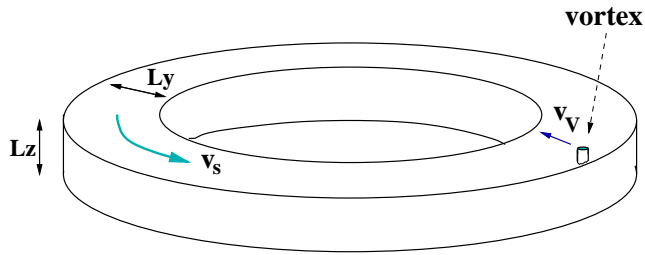}
\end{center}
\caption{Vortex being moved from the outside to the inside of a ring
  of superfluid, increasing the superfluid circulation around the ring
  by one quantum unit.}
\end{figure}

Consider a macroscopic ring, such as the one shown in fig.\ 1, with average
radius $R$, width (difference between outer radius and inner radius) $L_y$
and height $L_z$.  For simplicity we assume $L_y<<R$, but this is not
essential, and the result is independent of the shape of the ring.  Initially
there are $n$ quanta of circulation trapped in the ring, giving superfluid
velocity  $v_s=n\kappa_0/2\pi R$, and the normal fluid velocity is zero,
since the boundaries of the ring are stationary. A pinning potential is used
to insert adiabatically one vortex, which is created at the outer boundary,
moved slowly across the system under constant temperature conditions, and
annihilated at the inner boundary. The effect of this extra vortex is to
increase the circulation from $n$ units to $n+1$, increasing the superfluid
velocity by $\delta v_s=\kappa_0/2\pi R$. This increases the free energy by  
\be \Delta F= (2\pi RL_yL_z)\rho_s v_s\delta v_s=\rho_s\kappa_0
v_sL_yL_z\;,\ee
since superfluid density is defined in terms of the free energy change
when the superfluid velocity is changed.
This must be compared with the work done in moving the vortex
of length $L_z$ a distance $L_y$ isothermally across the ring, which is
\be W= F_tL_yL_z\;.\ee
Comparison of these two shows that the magnitude of the transverse force per
unit length, under conditions in which $v_n$ and $v_V$ are both negligible,
is
\be |F_t|=\rho_s\kappa_0 v_s\;.\ee
More careful analysis gives the sign and direction of this force as
\be {\bf F}_t= -\rho_s{\bf K}\times {\bf v}_s\;.\ee
This argument determines the coefficient of $v_s$ in the transverse force.

To determine the coefficient of $v_V$, Thouless, Ao and Niu
\cite{tan96} consider an infinite system with superfluid and normal
fluid asymptotically at rest (${\bf v}_n=0={\bf v}_s$) in the presence
of a single vortex which is constrained to move by moving the pinning
potential. For simplicity we describe the two-dimensional problem of a
vortex in a superfluid film, but the three-dimensional generalization
is straightforward.  Also we restrict this discussion to the ground
state of the vortex, but the generalization to a thermal equilibrium
state is straightforward. The reaction force on the pinning potential
is calculated to lowest order in the vortex velocity ${\bf v}_V$. This
can be studied as a time-dependent perturbation problem, but this can
be transformed into a steady state problem, with the perturbation due
to motion of the vortex written as $i{\bf v}_V\cdot{\bf grad}_0$.  The
force in the $y$ direction on a vortex moving with speed $v_V$ in the
$x$ direction can then be written as
\be F_y=iv_V\bigl< \Psi_{{\bf r}_0}\bigl|{\partial V\over\partial x_0} {{\cal
P}\over E_0-H}{\partial\over\partial y_0}\bigr|\Psi_{{\bf r}_0}\bigr>+\ {\rm
comp.\ conj.,} \ee
where $\cal P$ projects off the ground state of the vortex.   
Since ${\partial V/\partial x_0}$ is the commutator of $H$ with
the partial derivative $\partial/\partial x_0$, the denominator cancels with
the $H$ in the denominator, and so the expression is equal to the Berry phase
form  
\be F_y= -iv_V\bigl< {\partial\Psi\over\partial
x_0}|{\partial\Psi\over\partial y_0}\bigr>+iv_V\bigl<
{\partial\Psi\over\partial y_0}|{\partial\Psi\over\partial x_0}\bigr> \;.\ee

Since the Hamiltonian consists of kinetic energy, a translation invariant
interaction between the particles of the system, and the interaction with
the pinning center, which depends on the difference between the pinning
center coordinates and the particle coordinates, the derivatives
 $\partial/\partial x_0$, $\partial/\partial y_0$, can be replaced by the
total particle momentum operators $-\sum\partial/\partial x_j$,
$-\sum\partial/\partial y_j$. This gives the force as a commutator of
components, $P_x,P_y$ of the total momentum, 
\be F_y=-iv_V\bigl<\Psi_{{\bf r}_0}\bigl|[P_x,P_y]\bigr|\Psi_{{\bf r}_0}\bigr>
\;.\ee
At first sight one might think that the two different components of momentum
commute, but this depends on boundary conditions, since the momentum
operators are differential operators.  Actually this expression is the
integral of a curl, and can be evaluated by Stokes' theorem to get
\be F_y= \oint \bigl<\Psi_{{\bf
r}_0}\bigl|-i\sum_j{\bf grad}_j\bigr|\Psi_{{\bf r}_0}\bigr>\cdot\,d{\bf
r}=\oint {\bf j}_s\cdot d{\bf r}\;,\ee
where the integral is taken over a loop at a large distance from the vortex
core.  This gives the force in terms of the circulation of momentum density
(mass current density) at large distances from the vortex.

Our result that the transverse force is equal to $v_V$ times the line integral
of the mass current is independent of the nature or size of the pinning
potential. The general form of this is
\be {\bf F}_t= \rho_s{\bf K}_s\times {\bf v}_V +\rho_n{\bf K}_n\times {\bf
v}_V\;,\ee
where ${\bf K}_n$ represents the normal fluid circulation.

 In equilibrium the circulation of the normal fluid around a stationary
vortex is zero, since circulation of the normal fluid gives rise to viscous
dissipation of energy, which in turn leads to growth of the area of the
normal fluid vortex core. If there is any nonequilibrium normal fluid
circulation, it is not obvious that it should be quantized, or that the
motion of the normal fluid vortex should be correlated with the motion of
the superfluid vortex.  If only the superfluid participates in the circulation
round the vortex core, which seems to us to be the most reasonable
assumption, this gives 
\be {\bf F}_t= \rho_s{\bf K}\times {\bf v}_V\;.\ee 
In combination with the Wexler result for the coefficient of ${\bf v}_s$, the
total transverse force on a vortex is 
\be {\bf F}_t= \rho_s{\bf K}_s\times ({\bf v}_V-{\bf v_s}) \;.\ee 
Only the superfluid Magnus force exists unless the normal component
participates in the circulation of the superfluid to some extent.

This disagrees with Donnelly's eq.\ (\ref{eq:donnelly}) unless the
phonon-scattering term proportional to  $\sigma$ cancels with the Iordanskii
term proportional to $\rho_n$, and disagrees with Volovik's eq.\
(\ref{eq:volovik}) unless his coefficient $C_F$ is equal to $\rho_n$ even for
bosons.  The most striking feature of this is that the force is independent
of the normal fluid velocity.

It agrees with the obvious generalization of the classical Magnus
force argument to two-fluid dynamics.  This argument considers the
force-- momentum balance in a large cylinder surrounding a vortex which
is held stationary while the fluid flows past it.  Bernoulli pressure
on the cylinder, and momentum flux across the boundary of the cylinder
balance with the force on the vortex.  In a two-fluid generalization
of this there are separate contributions from the product of
superfluid circulation with superfluid velocity, and from the product
of normal fluid velocity with normal fluid velocity.

Since we have only had to consider global properties involving momentum
conservation and conditions at a long distance from the vortex core, and
have not needed to make any detailed consideration of conditions at the
core of the vortex, we believe that our arguments are valid for a fermion
superfluid described by a single complex order parameter.  The $B$ phase of
$^3$He does not quite meet this condition, but the order parameter is an
essentially isotropic combination of a $P$-wave orbital state and a triplet
spin, so this should behave in much the same way.  Volovik
\cite{volovik93,volovik95} has argued that spectral flow of the unpaired
states in the vortex core of a fermion superfluid leads to a contribution to
the transverse force that cancels most of the Magnus force, but Stone
\cite{stone96} has examined this argument more closely, and does not find that
this mechanism is operative unless there is a background to take momentum from
these excitations.  We do not think that there is such a canceling
contribution in a homogeneous fermion superfluid.  \vskip 18pt

\section{Forces due to phonon scattering}

The result obtained in the previous section, that the coefficient of the
vortex velocity in the transverse force is equal and opposite to the
coefficient of the superfluid velocity leads to the surprising conclusion
that normal fluid flow does not affect the force on the vortex, unless there
is also normal fluid circulation round the vortex.  This is surprising,
because Pitaevskii \cite{pitaevskii??} and Iordanskii \cite{iordanskii66}
argued that the asymmetrical scattering of rotons or phonons by vortices
should lead to a transverse force when the vortex moves relative to the
normal fluid component.  

In the low temperature limit an explicit calculation of the phonon--vortex
scattering can be made, and the literature quotes a transverse force
proportional to $\rho_n{\bf K}\times ({\bf v}_V-{\bf v}_n)$.  There are two
problems with this result:
\begin{enumerate}
\item The derivation assumes that the phonons interact only with the
vortex, but in our argument we assume that the phonons, which make up the
normal fluid, must be in equilibrium with one another.

\item In papers from Cleary (1968) \cite{cleary68} to Sonin (1997)
\cite{sonin97} the expression for the transverse force, which is proportional
to 
\be 4{T^3\over \hbar^2 c^3}( v_V- v_n)\sum_m\sin \delta_m\sin\delta_{m+1}
\sin(\delta_{m+1}-\delta_m)\;,\ee  
has been rewritten as
\be {T^3\over \hbar^2 c^3}( v_V- v_n)\sum_m\sin
2(\delta_{m+1}-\delta_m)\;.\ee 
This would be fine, except that $\delta_m$
does not tend to zero.  If one substitutes the formula
\be \delta_m\propto {\rm sign}(m)\kappa_0 T/\hbar c^2\;,\ee
which is correct to lowest order in temperature $T$, into the original
formula, a result is obtained which is (at least) cubic in $\kappa_0$ and of
sixth power in $T$, or 3/2 power in $\rho_n$.  The second expression is
obtained from the first by canceling two divergent series, and this gives
the quoted expression which is linear in $\kappa_0$ and linear in $\rho_n$;
we cannot see any justification for a term of this magnitude. 
\end{enumerate}\vskip 18pt

\section{Superconductivity}

The situation for the transverse force on a vortex in a superconductor is
even more confused than the situation for a neutral superfluid.  In the
1960s, Bardeen and Stephen \cite{bardeen65} argued for a very small Magnus
force, but an analysis by Nozi\`eres and Vinen \cite{nozieres66} of an
idealized model of a superconductor gave the full value of the Magnus force
suggested by classical hydrodynamics. 

Wexler's argument \cite{wexler97} for the coefficient of $v_s$ can be
applied to the case of a superconductor.  When the substrate velocity
and the vortex velocity zero in the presence of a superfluid electron
velocity, this argument gives the expected result that there is a
Lorentz force on the vortex equal to the integral of $e\rho_e{\bf
v}_s\times{\bf B}$, where $\rho_e$ is the conduction electron density.

To find the coefficient of $v_V$, Geller, Wexler and Thouless \cite{gwt98} have
adapted the arguments of sec.\ 4 to the very idealized model of a charged
system with a uniform positive background, which is essentially the situation
considered by Nozi\`eres and Vinen \cite{nozieres66}, although they, unlike us,
also had to assume that the superconductor was extreme type II.  This is not
completely straightforward, even though we have taken the uniform positive
background so that we can continue to use momentum conservation, because any
choice of the gauge field which is used to describe magnetic effects breaks
the explicit translation invariance, and makes the implicit translation
invariance obscure. Rather than introduce a gauge field, we can write the
electromagnetic interactions in terms of a Coulomb interaction between
electrons and between electrons and positive background, together with an
instantaneous current--current interaction between the electrons.  Darwin
showed that this is correct up to second order in electron velocity, apart
from a relativistic variation of the mass with velocity which is unimportant
for this problem.  We also need a Galilean invariant attractive interaction
between the charges to produce a paired superconducting state.  This gives a
Hamiltonian with explicit translation invariance, so the arguments of sec.\ 4
can be taken over.

The result for the coefficient of $v_V$ is formally unchanged, and is the line
integral of the canonical momentum density on a loop which surrounds the flux
line at a distance which is large compared with the penetration length. 
This is actually a surprising result, as at these distances there is
no magnetic field or current density produced by the vortex line, and the
integral is related to the Aharonov--Bohm effect rather than to any
classical quantity. 

Since the integral is equal to the trapped magnetic flux, the
transverse force can be written as
\be {\bf F}_t= \int\!\!\!\int\!\!\!\int \rho_e ({\bf v}_V-{\bf v}_e)\times
{\bf B}\,d^3r\;,\ee
so that the transverse force depends only on the motion of the vortex
relative to the electrons.  It can be rewritten, in a form that makes its
physical origin more transparent, as 
\be {\bf F}_t=\int\!\!\!\int\!\!\!\int \rho_e ({\bf v}_p-{\bf
v}_e)\times {\bf B}\,d^3r +\int\!\!\!\int\!\!\!\int \rho_e ({\bf v}_V-{\bf
v}_p)\times {\bf B}\,d^3r\;.\ee 
The first term is the Lorentz force given by the interaction of the
electric current, which is a Galilean invariant, with the magnetic field. 
The second is a Magnus force that acts on the positive substrate moving with
velocity ${\bf v}_p$ relative to the vortex.  The moving vortex generates a
dipolar electric charge distribution, which in turn produces a dipolar
elastic stress on the positive substrate, and this leads to a net force on
the positive substrate. 

A similar analysis was carried out by Nozi\`eres and Vinen \cite{nozieres66},
and their results were essentially the same.  There have been recent
measurements made by Zhu, Brandstrom and Sundqvist \cite{zhu97} that support
a fairly large value of the Magnus force.
 
\section{Conclusions}

We have succeeded in determining the transverse force on a vortex in a
neutral superfluid under assumptions that are both general and reasonably
realistic.  Like most other exact results in quantum many-body theory, they
are related to general conservation laws, and apply, in a slightly different
form, to classical systems as well.

Our generalization to superconductors is far from realistic, since it relies
on the uniformity of the positive substrate.  With a uniform positive
substrate an electron gas has infinite conductivity, even in the absence of
a pairing interaction, so our results can only form a first step towards a
plausible theory of the Magnus force in superconductors.  We may be able to
extend the reults from a uniform substrate to an ideal periodic sustrate,
but even that is quite inadequate for the description of a real metal.  We
have to be able to take the next step of considering disorder, but there is
no chance that we will be able to get exact results in that case.

It would also be interesting to generalize these results to finite systems,
nonzero frequency of the vortex motion, and a finite density of vortex lines.

Another line that we are pursuing is the connection between the Magnus force
and the quantization of the vortex line.  We know that there is an intimate
connection between the strength of the Lorentz force on an electron and the
density of degenerate levels of an electron in a magnetic field, and there
are good reasons to think that there is a similar connection between the
strength of the Magnus force on a vortex and the density of degenerate
levels for a vortex.

\section*{Acknowledgments}

This work was supported in part by NSF Grant No.\ DMR 95-28345.


\end{document}